\documentclass[a4paper]{spie}  %>>> use for US letter paper
\usepackage{aas_macros}
\usepackage[]{graphicx}
\usepackage{eurosym}
\usepackage{gensymb}
\usepackage{url}

\def\arcsec{\hbox{$^{\prime\prime}$}}

\title{ExTrA: Exoplanets in Transit and their Atmospheres} 

\author{X.~Bonfils, J.M.~Almenara, L.~Jocou, A.~Wunsche, P.~Kern, A.~Delboulb\'e, X.~Delfosse, P.~Feautrier, T.~Forveille, L.~Gluck, S.~Lafrasse, Y.~Magnard, D.~Maurel, T.~Moulin, F.~Murgas, P.~Rabou, S.~Rochat, A.~Roux, and E.~Stadler
\skiplinehalf
Univ. Grenoble Alpes, IPAG, F-38000 Grenoble, France\\
CNRS, IPAG, F-38000 Grenoble, France\\
}

\authorinfo{Further author information: (Send correspondence to X.B. \& J.M.A.)\\ E-mail: xavier.bonfils@obs.ujf-grenoble.fr, jose-manuel.almenara-villa@ujf-grenoble.fr}
\begin{document} 
  \maketitle 

%%%%%%%%%%%%%%%%%%%%%%%%%%%%%%%%%%%%%%%%%%%%%%%%%%%%%%%%%%%%% 
\begin{abstract}
The ExTrA facility, located at La Silla observatory, will consist of a near-infrared multi-object spectrograph fed by three 60-cm telescopes. ExTrA will add the spectroscopic resolution to the traditional differential photometry method. This shall enable the fine correction of color-dependent systematics that would otherwise hinder ground-based observations. With both this novel method and an infrared-enabled efficiency, ExTrA aims to find transiting telluric planets orbiting in the habitable zone of bright nearby M~dwarfs. It shall have the versatility to do so by running its own independent survey and also by concurrently following-up on the space candidates unveiled by K2 and TESS. The exoplanets detected by ExTrA will be amenable to atmospheric characterisation with VLTs, JWST, and ELTs and could give our first peek into an exo-life laboratory.
\end{abstract}

%>>>> Include a list of keywords after the abstract 


%%%%%%%%%%%%%%%%%%%%%%%%%%%%%%%%%%%%%%%%%%%%%%%%%%%%%%%%%%%%%
\section{INTRODUCTION}
\label{sec:intro}  % \label{} allows reference to this section

We are living remarkable times of astronomy. Since the ground-breaking discoveries of other planets outside our Solar System \cite{1992Natur.355..145W,1995Natur.378..355M,1996ApJ...464L.147M}, the path towards the detection of other Earths, towards their characterisation and towards the possible evidences of Life elsewhere in the Universe is unfolding before us. Super-Earths with masses below 10~M$_{\oplus}$ and orbiting in the ``Habitable Zone'' of their stars are being found\cite{2007A&A...469L..43U,2009A&A...507..487M, 2011A&A...534A..58P, 2013A&A...549A.109B,2013A&A...553A...8D, 2013A&A...556A.110B, 2012ApJ...745..120B}. For a range of plausible albedos and atmospheric compositions, liquid water may flow on their surface\cite{1993Icar..101..108K,2007A&A...476.1373S} and, since liquid water is thought as a prerequisite for the emergence of life as we know it, these planets constitute a prized sample for further characterisation of their atmosphere and the search for possible bio-signatures.

The spectroscopy of exoplanets atmospheres is much easier for transiting planets than non-transiting planets and much interest is therefore focused on finding Earth-like planets that transit stars bright enough for detailed spectroscopic follow-ups. When a transiting planet passes in front of the stellar disk, its opaque body blocks a fraction of the stellar light, and its atmosphere subtly filters an even smaller fraction. Chromatic measurements of transit light curves thus permit to retrieve the transmitted spectra and to infer the atmospheric opacity sources\cite{2000ApJ...537..916S}. This was accomplished for the first time using HST to detect neutral sodium in the atmosphere of HD~209458b\cite{2002ApJ...568..377C} and later confirmed with ground-based observations\cite{2008A&A...487..357S}. It has been followed since by numerous detections of other species in either HD~209458b' or HD~189733b's atmospheres, like H2, H2O, CH4, TiO and CO (e.g., Ref.~\citenum{2007Natur.448..169T} and Ref.~\citenum{2008A&A...492..585D}); although many remain controversial (see review in Ref.~\citenum{2010ARA&A..48..631S}). Transmission spectroscopy has also revealed atmospheric escape\cite{2003Natur.422..143V}, high altitudes hazes\cite{2008MNRAS.385..109P}, and winds\cite{2010Natur.465.1049S}. The atmospheres of exoplanets can also be studied by the technique of occultation spectroscopy. The star-only spectrum is observed when the planet passes behind the star and is subtracted to the combined star+planet spectrum observed just before or after the occultation. Thermal emission from hot-Jupiters were first measured this way with Spitzer\cite{2005Natur.434..740D,2005ApJ...626..523C} and since made with several other telescopes from ground and space. Occultation spectra are also modulated by the temperature-pressure profile, the chemical composition and the climate of the planetary atmospheres (e.g. Ref.~\citenum{2007Natur.447..183K}).

Both transmission and occultation spectroscopy require extreme signal-to-noise ratios (S/Ns). Per unit of observing time and for a given planetary size, better S/Ns are achieved with brighter stars. And whereas the wealth of small exoplanets detected by the space missions CoRoT and especially Kepler has provided a new statistical perspective on planetary systems, these planets orbit stars too faint for detailed spectroscopic follow-up. As a consequence, the new space missions and ground based facilities that aim to detect exoplanets more amenable to atmospheric characterisation are now focused on nearby, brighter stars.

For a given planetary size, better S/Ns are also achieved with smaller stars. Small planets transiting bright and late-type stars, such as nearby M~dwarfs, thus have great added value. So much that habitable Earth-like planets transiting M~dwarfs, aided by transmission and occultation spectroscopy, are contemplated as the shortest route to peek into an exo-Life laboratory (e.g. Ref.~\citenum{2009ApJ...698..519K, 2013ApJ...764..182S}). And, before being privileged targets for the forward characterisation of their transiting planets, M~dwarfs are, in the first place, easier targets for the detection of these planets. An Earth-size planet transiting a 0.1-R$_\odot$ M dwarf eclipses 0.8\% of the stellar surface. This is a hundred times more than if the same planet were to transit a Sun-like star. The proximity of the habitable zone to the star is also an advantage for habitable planets, which are 4-10 times more likely to transit if they orbit an M instead of a G dwarf. Planets at a given separation produce larger amplitude radial-velocity variations around lower-mass stars, which eases the mass measurement of planets detected in transit. Last but not least, we already know that super-Earths in the habitable zone of M~dwarfs are very frequent\cite{2013A&A...549A.109B, 2015ApJ...807...45D}.

In that context, several strategies are pursued to search for planets transiting bright M~dwarfs. From the ground, they can only be observed one at a time, either with a dedicated photometric search (e.g. GJ1214b by the MEarth survey\cite{2009Natur.462..891C}), or with a Doppler spectrograph (e.g. GJ3470b detected by the HARPS radial-velocity survey and a focused photometric follow-up\cite{2012A&A...546A..27B}). From space, K2 and TESS are foreseen to detect planets as small as 2~R$_\oplus$ planets in the habitable zone of bright M~dwarfs\cite{2014arXiv1411.1754B, 2015arXiv150603845S}.

\begin{figure}
  \begin{center}
    \begin{tabular}{c}
      \includegraphics[height=8cm]{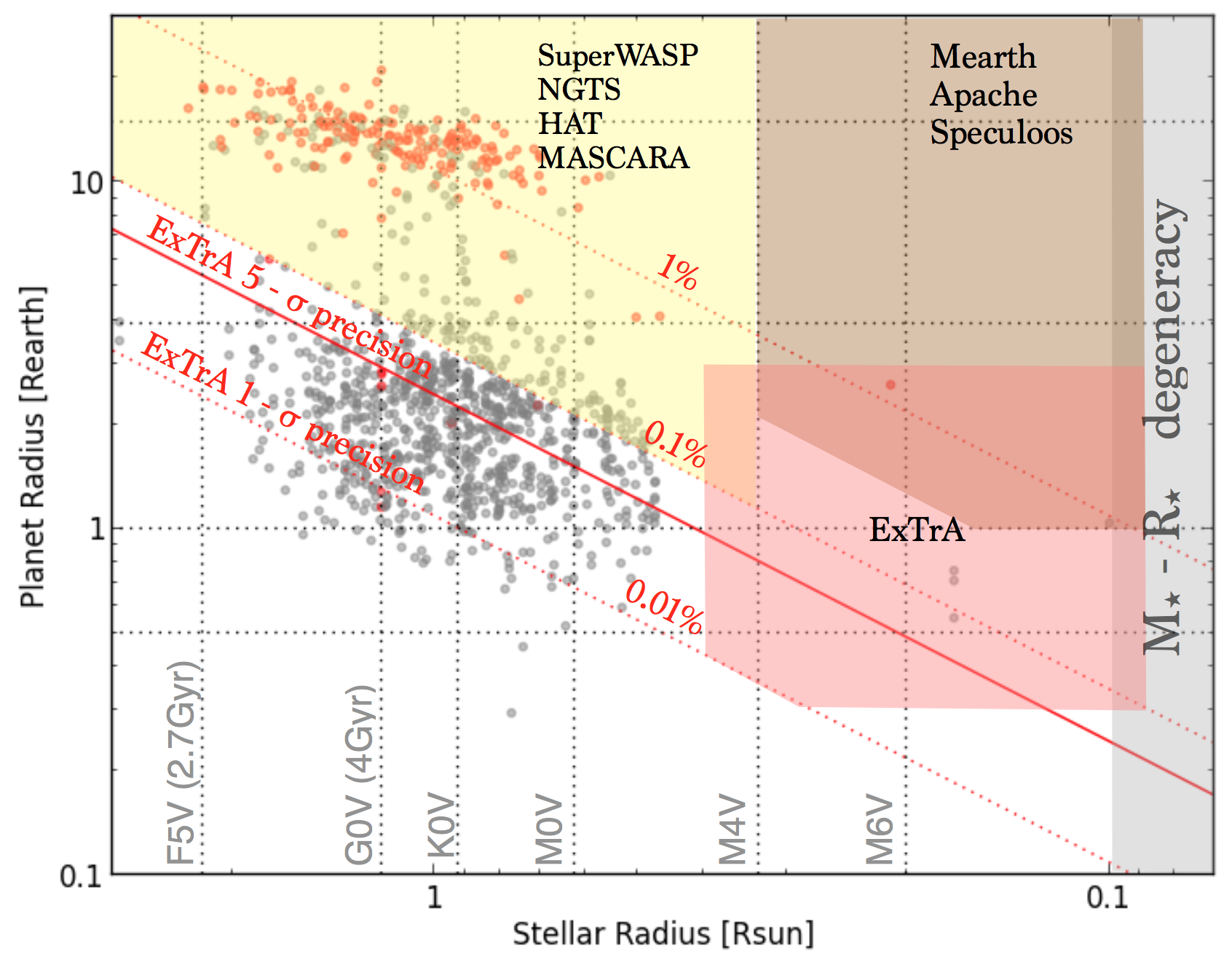}
    \end{tabular}
  \end{center}
  \caption[Parameter space of transit suveys]{Parameter space of selected transit suveys. Wide-field surveys targeting thousands stars per field with \textless20~cm telescopes detect planets in the yellow zone. Targeted surveys observing M~dwarfs one-by-one with 40-80~cm telescopes are sensitive to planets in the orange zone. Known transiting planets are reported, in red if detected by ground-based surveys and in gray if detected by CoRoT and Kepler. Space detections by CoRoT and Kepler are around stars too large and too faint for spectroscopic characterization. Transit depths of 1\%, 0.1\% and 0.01\% are shown with dotted red lines. The red zone shows the parameter space covered by ExTrA.\label{fig:spaceparam}}
\end{figure}

Figure~\ref{fig:spaceparam} shows the parameter space covered by both wide-field surveys (superWASP, HAT, NGTS, MASCARA and alike) and targeted surveys (MEarth, Apache, and alike). Wide-field surveys (yellow zone) are sensitive to planets around F-to-early M~stars, with G and K~dwarfs being the bulk of their targets. Targeted surveys (orange zone) are designed to search for planets around late-M~stars. Both approaches are progressively reaching the 1 Earth radii sensitivity. The figure also shows iso-contours for the corresponding transit depths (red dotted lines).

The project \emph{Exoplanets in Transit and their Atmosphere} (ExTrA), aims to find transiting telluric planets orbiting in the habitable zone of bright nearby M~dwarfs. Such exoplanets will be amenable to a follow-up characterization with VLTs, JWST, and ELTs and will give our first peek into an exo-life laboratory. In particular, we developed a novel method sensitive to \textgreater0.5~R$_{\oplus}$ planets (Fig.~\ref{fig:spaceparam}, red zone) that we propose to apply by building and operating the ExTrA facility (three 60-cm telescopes coupled to a single infrared multi-object spectrograph). This novel method overcomes the limitations of previous transit searches. It aims at a photometric precision of 0.02\% per measurement, usually only achieved in space. Being in the infrared, it is also far more efficient in observing late-type M~dwarfs.

This document presents the project. We present the novel method proposed for ExTrA (Sec.~\ref{sec:extra}). We describe its implementation and show it is an order of magnitude more precise (Sec.~\ref{sec:performances}) and more than an order of magnitude more efficient than traditional ground-based photometry. We outline our survey for a nominal duration of 4~years (Sec.~\ref{sec:survey}).

ExTrA was selected by the European Research Council and was awarded the maximal grant of 2M\euro\ to fund the project during its first 5~years (2014-2019).

\section{ExTrA}
\label{sec:extra}

\subsection{A novel method}

From the ground, photometric light curves are limited by systematics due to changing air mass, atmospheric conditions, telescope pointing or flat-field errors, that result in a correlated ``red'' noise\cite{2006MNRAS.373..231P}. So far, this has prevented ground-based surveys to search for \textless 2 R$_{\oplus}$ planets.
In response, we have developed a novel method that aims to overcome previous limitations and push down the precision of ground-based differential photometry. Our project will use a near-infrared spectrograph fiber-fed by three 60-cm telescopes. On each telescope, 5 Field Units (FUs) will be used in place of the CCD camera to pick off the light from the target and selected comparison stars. The (spectro-)photometry of the target will be measured differentially with the (spectro-)photometry of the 4 comparison stars. To ensure photometric stability, the FUs will be composed by fibers accurately positioned. The facility will count 3 independent telescopes and all fibers from all 3 telescopes will feed a single near-infrared spectrograph that has a low spectral resolution (R$\sim$200) and that covers the 0.9-1.6~$\mu$m range. Although all three telescopes can observe the same field, they will most of the time observe different fields so to maximize the transit search efficiency.

This experimental setup presents many advantages compared to current ground-based photometric searches:
\begin{itemize}
\item Atmospheric absorption: With traditional differential photometry, the spectra of comparison stars never match perfectly the spectrum of the target. When the atmospheric absorption is chromatic, the amount of light that is filtered out by the Earth atmosphere is slightly different for the comparison stars and for the target. Ground-based transit searches thus have to select a narrow band-pass with as few telluric lines as possible. For instance, when the 60-cm telescope TRAPPIST observed the transit of GJ3470b\cite{2012A&A...546A..27B} we used a z'-Gunn filter to improve on precision. Combined to the poor quantum efficiency of the CCD beyond 900~nm, this resulted in a band pass of only $\sim$60~nm centered on 860~nm. With traditional differential photometry, better precision implies to use telescopes always larger with filters always narrower.
ExTrA uses a (multi-object) spectrograph that resolves 200 spectral channels of \textless 6~nm each (R$\sim$200). On the one hand, the resolution enables differential photometry with much narrower spectral channels, and thus with much finer correction of the atmospheric variations. On the other hand, the wide spectral domain (0.9-1.6~$\mu$m) enables the simultaneous collection of more photons, and thus an increased efficiency. Once the differential photometry is done on individual spectral channels, we will thus degrade the spectral resolution to produce broad band light curves with high signal-to-noise ratios.

\item Wavelength domain: Ground-based transit searches usually require detectors with large areas (e.g. 2k~x~2k format). In the optical, cameras with large CCD sensors are available at an entry price low enough to equip many telescopes. Although there are now efforts to use low-cost InGaAs detectors\cite{2013PASP..125.1021S, 2014SPIE.9154E..1FS}, large infrared detectors remain very expensive. For example, the HAWAI2RG detector developed by Teledyne is commercialized for about \$250,000, and the cameras integrating this chip have an entry price above \$500,000. 

ExTrA, however, has the advantage of multiplexing: we can use a single spectrograph (and thus a single camera) to record the light from several telescopes. The spectrograph will also use the detector area much more efficiently than imagers and can thus accommodate to smaller detectors, with a camera cost below \$100,000. This presents a unique opportunity to observe at infrared wavelengths at a moderate cost, which is an enormous advantage when observing M dwarfs. Figure~\ref{fig:sed} shows the photon distribution for a M5, a M9, and a L3 dwarfs. They respectively produce 9, 22, and 40 times more photons between 0.9-1.6μm than between 0.7-0.9~$\mu$m.

\begin{figure}
  \begin{center}
    \begin{tabular}{c}
      \includegraphics[height=6cm]{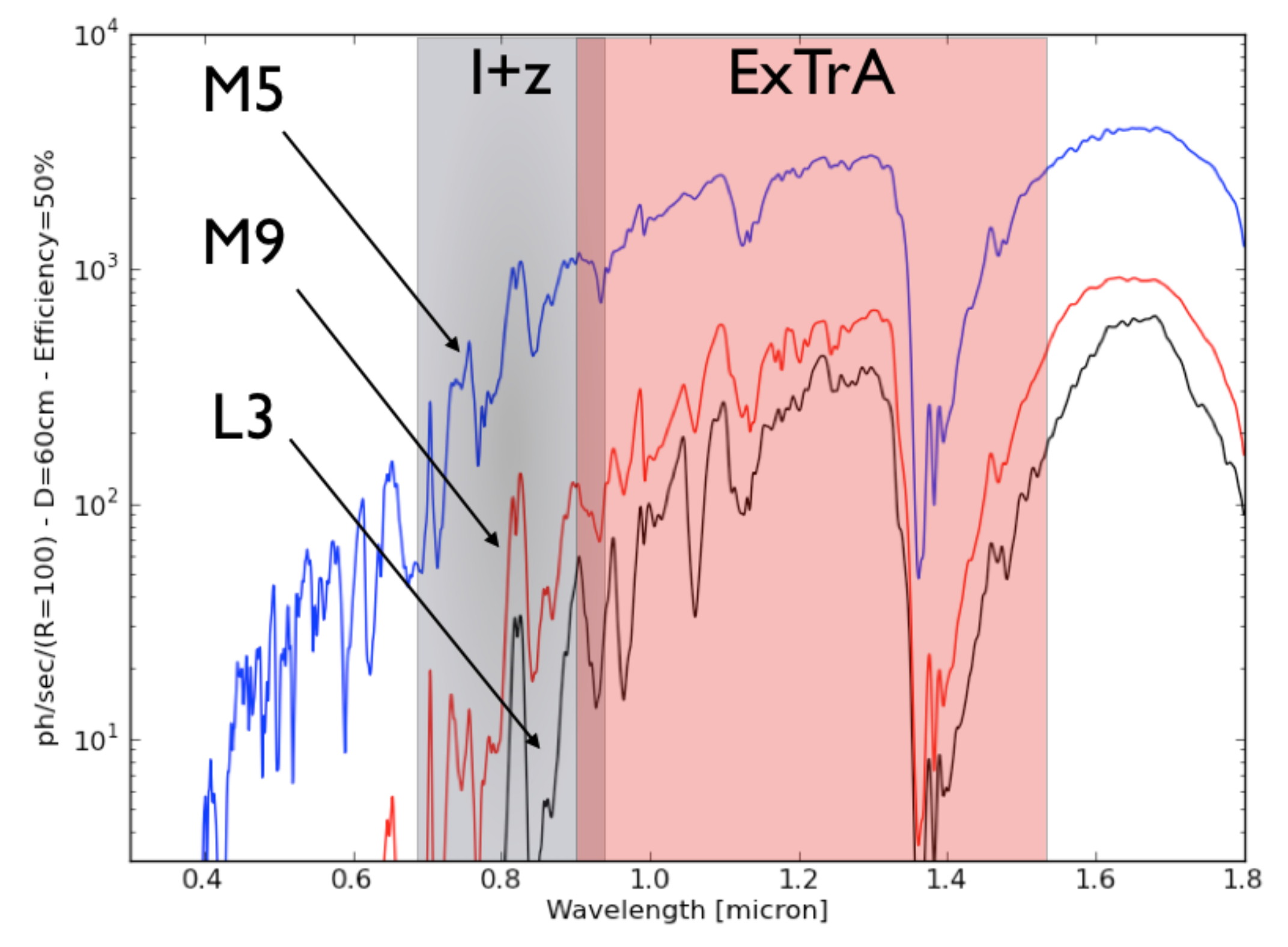}
    \end{tabular}
  \end{center}
  \caption[Parameter space of transit suveys]{Photon distribution (per R=100 resolution element) for a M5@15pc, a M9@10pc, and a L3@10pc dwarfs, as would be recorded with a telescope of 60-cm diameter, with a 50\% efficiency, and through the Earth atmosphere. The gray zone labeled €˜'I+z'€™ is the typical band pass of ground-based transit surveys. The light-red zone labeled 'ExTrA' is our spectral window. Stellar spectra are derived from NextGen models\cite{1997ARA&A..35..137A} and the Earth atmospheric transmission from ATRAN\cite{1992Lord}.\label{fig:sed}}
\end{figure}

\item Detector systematics, seeing variations and guiding errors: Detector pixels also have their imperfections. Their non-linearity, intra-pixel response, respiration, charge transfer efficiency and remanence can limit their accuracy to about 0.1\%. A good practice of differential photometry is to always guide on the same pixel. Although it mitigates some systematics, seeing variations and guiding errors remain and make impossible to maintain an identical illumination for a given pixel. An additional common practice is thus to also defocus the telescope: the comparison stars and the target point spread functions (PSFs) are then sampled by many more pixels (up to a few hundred), and the systematics average down with the square root of their number.
With ExTrA, each spectra will be dispersed on $\sim$3000 pixels, hence ensuring excellent averaging of detector systematics. The chosen FU design also permit excellent centering, even when mechanical flexures from the telescope or imperfect polar alignment distort the field geometry.
\end{itemize}

The concept we have developed overcomes the limitations of previous photometric surveys. Below, we also show it has the sensitivity and (photometric) stability to achieve a precision of $\sim$0.02\%.hr$^{-1/2}$ for stars brighter than J=11 mag.

\subsection{Design}

\subsubsection{Telescopes}

All elements were chosen based on a cost/benefit analysis. The cost of small telescopes scales approximately with the surface of the primary mirror and a telescope of diameter D is therefore as efficient as 2 telescopes of diameter D/$\sqrt{2}$. Using many smaller telescopes is therefore tempting, as this will average down some systematics. Such a setting however requires additional Field Units (see Sec.~\ref{sec:fieldunits}) and therefore ends up being significantly more expensive. As long as the cost scales as D$^2$, larger telescope diameters are therefore more cost-effective. A breakpoint however appears at D=60~cm, above which telescopes need larger and more robust mounts, which are far more expensive. ExTrA will use three 60~cm Ritchey-Chretien telescopes provided by AstroSystem Austria (ASA). We need a large field of view of 1\degree\ to have enough convenient comparison stars. This requirement give an aperture number of F/8 with a 60~cm diameter of primary mirror. The plate scale is 23~$\mu$m/\arcsec. A 2-lens corrector, optimized for zero field curvature, provides a telecentricity less than 21~mrd (an important aspect for the light injection in the fibers).

We choose an equatorial mount by having a non-rotating field of view. Direct drive technology allows us to have no backlash compensation and good performance in terms of tracking. The DDM160 mount, also provided by ASA, has a pointing precision of 6\arcsec\ RMS, and 0.25\arcsec\ RMS guiding precision without autoguiding. Moreover, we will use a guiding camera to get rid of the pointing precision and reduce the guiding error to \textless 0.05\arcsec. The guiding camera will be fixed in an off-axis position, this implies that we need a large sensor chip (10' square field of view) to assure an appropriate guiding star (V\textless13) in the ExTrA fields. This specifications are met by the Trius SX16 of Starlight Xpress.

\begin{figure}
  \begin{center}
    \begin{tabular}{c}
      \includegraphics[height=8cm]{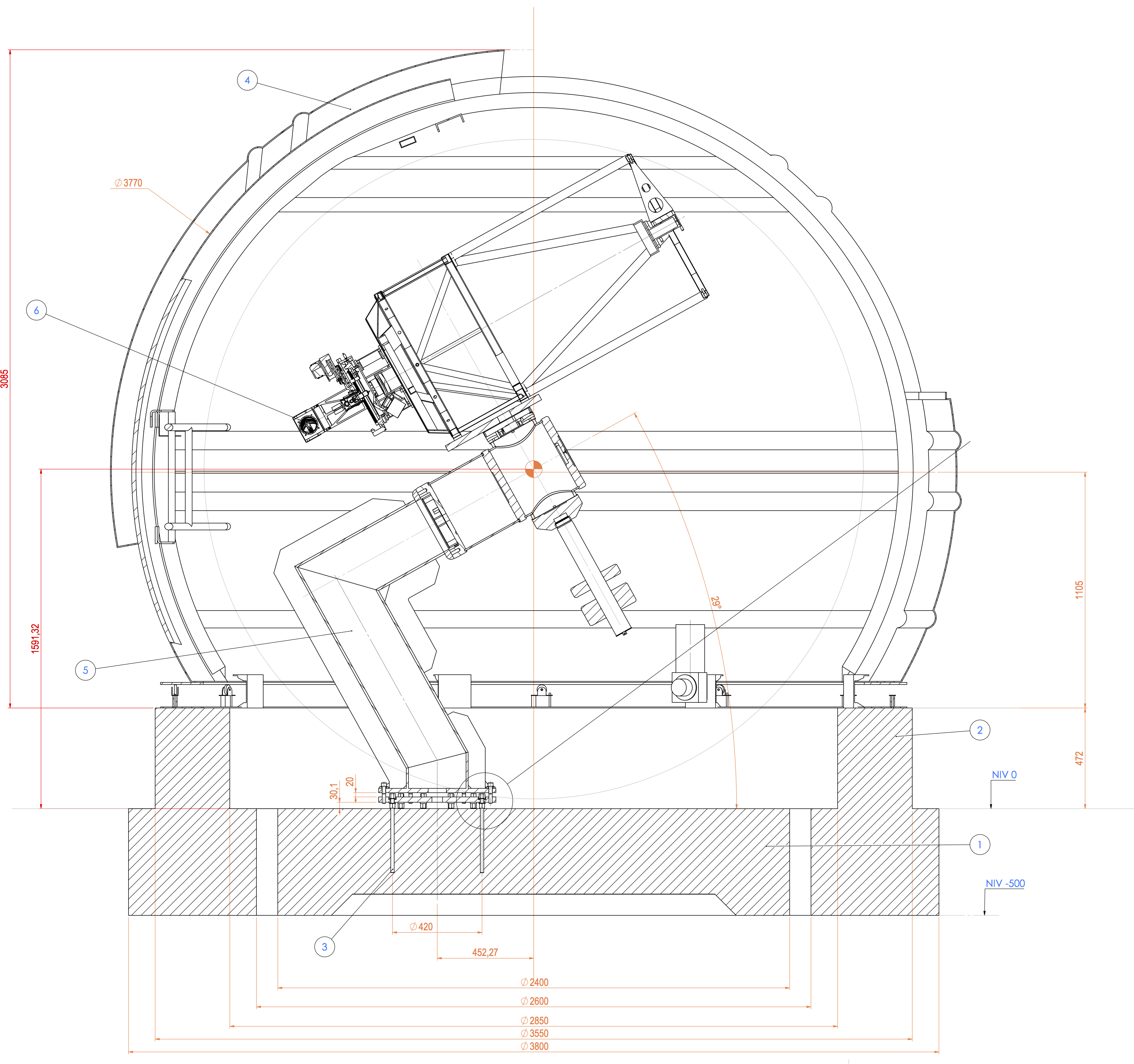}\includegraphics[height=8cm]{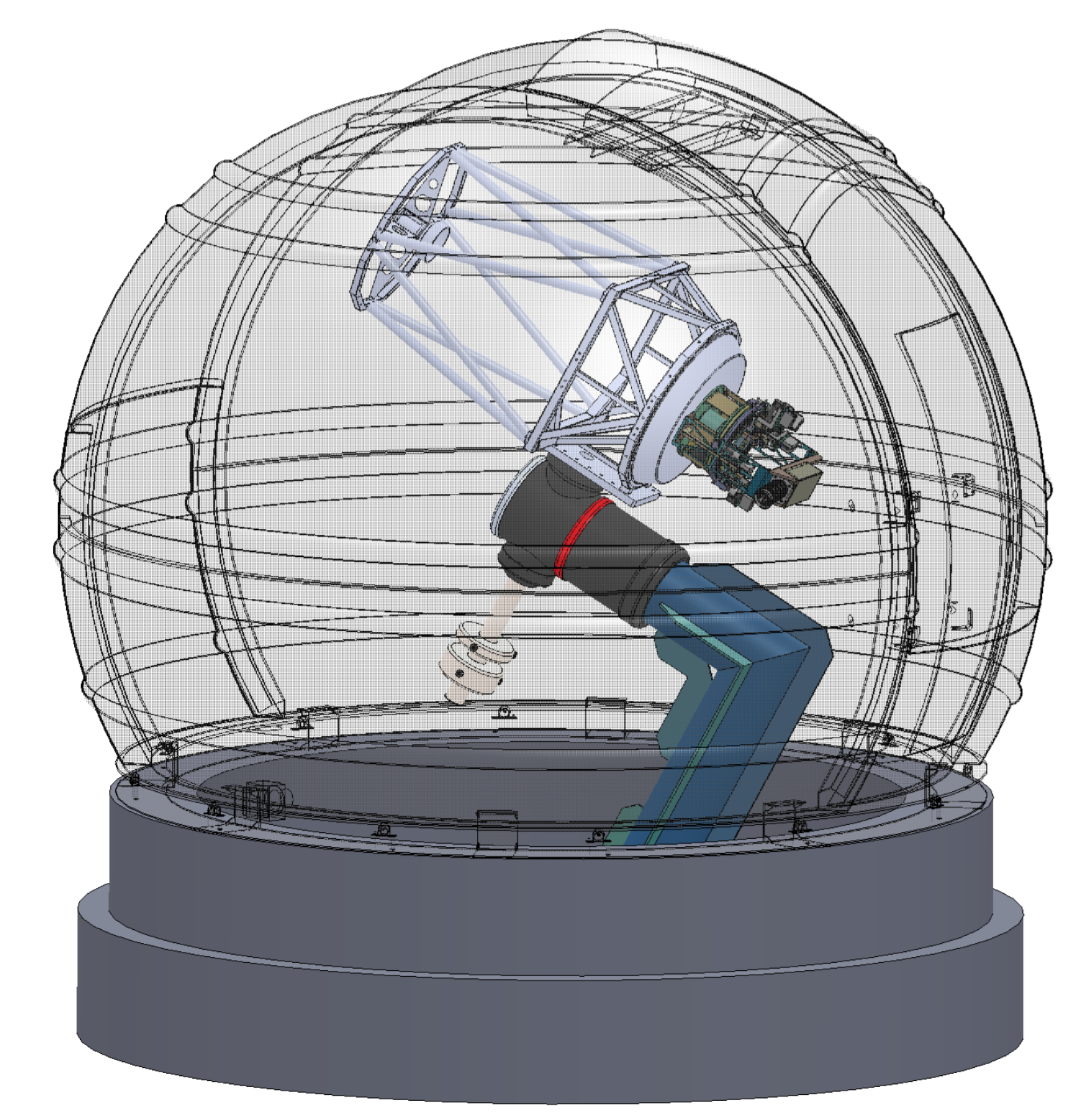}
    \end{tabular}
  \end{center}
  \caption[Telescope]{Plan (left panel) and 3D implementation (right panel) of one of the three ExTrA domes with the RC~600~ASA telescope, the DDM160~mount, and the focal plane unit.\label{fig:telescope}}
\end{figure}

Each of the three telescope is enclosed in a 4-m diameter dome made by Scopedome (see Fig.~\ref{fig:telescope}). The dome and the mount pier lie on separate concrete foundations to avoid any vibration disturbance.

\subsubsection{Field units}
\label{sec:fieldunits}

The Field Units (FUs) are designed to collect the F/8 telescope beam of the target and comparison stars and inject it into the optical fibers. Each FU is composed of two ``buttons'': the science button, and the centering button. Due to the error in absolute positioning of the FUs, the centering button is composed by a bundle of 37 fibres, used to calculate the position of the stars accurately, and then apply a relative movement (precise to 1~$\mu$m) to center the star in the science button.

Each science button is composed by two channels, one optimized for bright stars with an aperture of 8\arcsec\ and a second one optimized for faint stars with an aperture of 4\arcsec\ (to reduce the background contribution). Each channel have two fibers: one to collect the light of the star and another to collect the sky background close to the star.

The direct injection of star images into fibers is known to have some undesirable effects. The fiber conserve the spatial information between the entrance and the exit, thus making the system sensitive to seeing fluctuations and image centering. To get around this effects, we decided to use an aspherical microlens array in front of the fiber to image the pupil of the telescope on the fiber entrance. In this configuration, the atmospheric turbulence and guiding errors have a limited impact on the fiber injection, and as a result the flux distribution at the fiber exit is more stable.

The microlens are chrome masked to limit the aperture to 8\arcsec\ or 4\arcsec. We have choose circular fibers with a diameter of 150~$\mu$m due to their impact on the spectrograph resolution. To maximize the transmission of the fiber, the cladding thickness must be $\textgreater 15\ \lambda$.

\begin{figure}
  \begin{center}
    \begin{tabular}{c}
      \hspace{-0.2cm}\includegraphics[height=7.8cm]{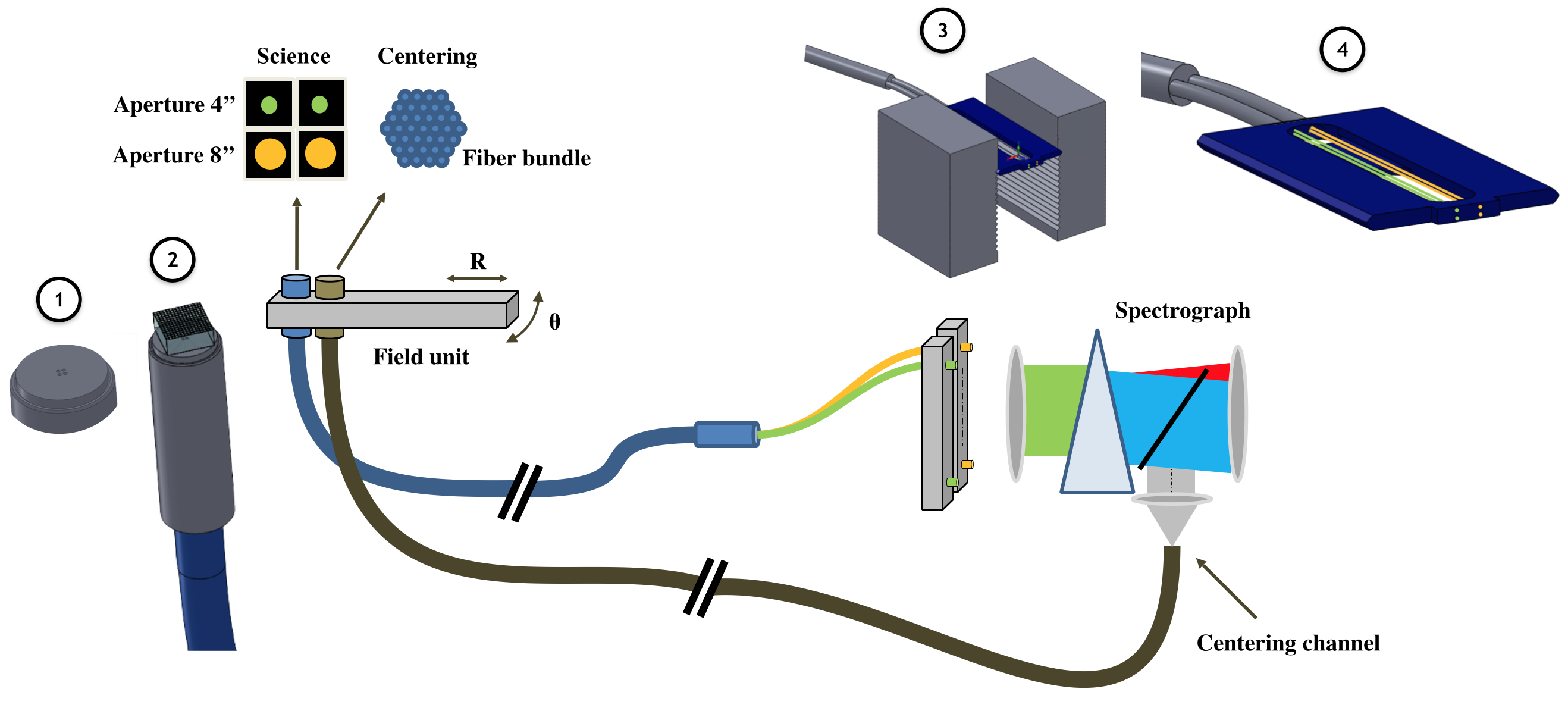}
    \end{tabular}
  \end{center}
  \caption[Fibers]{Scheme of the ExTrA fiber system from the field units to the spectrograph, with the 3D implementation of some of the parts: detail of a science button (1), science button with microlens (2), drawner arrangement at the spectrograph image plane (3), and drawner detail (4).\label{fig:fibers}}
\end{figure}

Each telescope have 5 FUs, each with 4 science fibers and one fiber bundle, thus 60 fibers and 15 bundles, of 20 meters length, connect the focal plane of the telescopes with the spectrograph (see Fig.~\ref{fig:fibers}).
The system is designed for an easy fiber set replacement. 

In ongoing lab tests of these FUs, we already achieved a photometric precision of $\sim$0.01\%, providing the position of the star is maintained to $\pm 2~\mu m$. 

\subsubsection{Focal plane positionners}

\begin{figure}
  \begin{center}
    \begin{tabular}{c}
      \includegraphics[height=8cm]{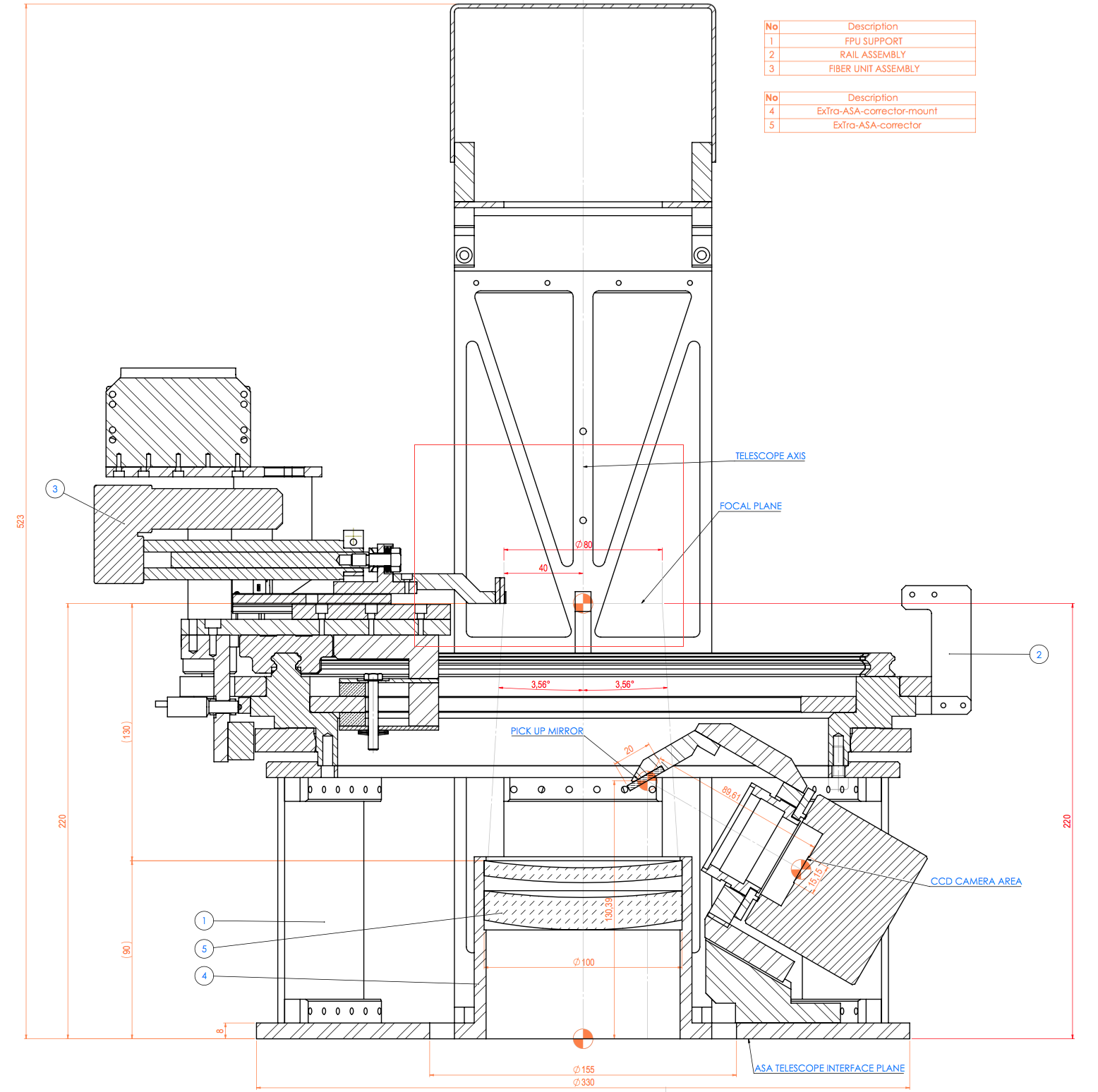}\hspace{-0.1cm}\includegraphics[height=8cm]{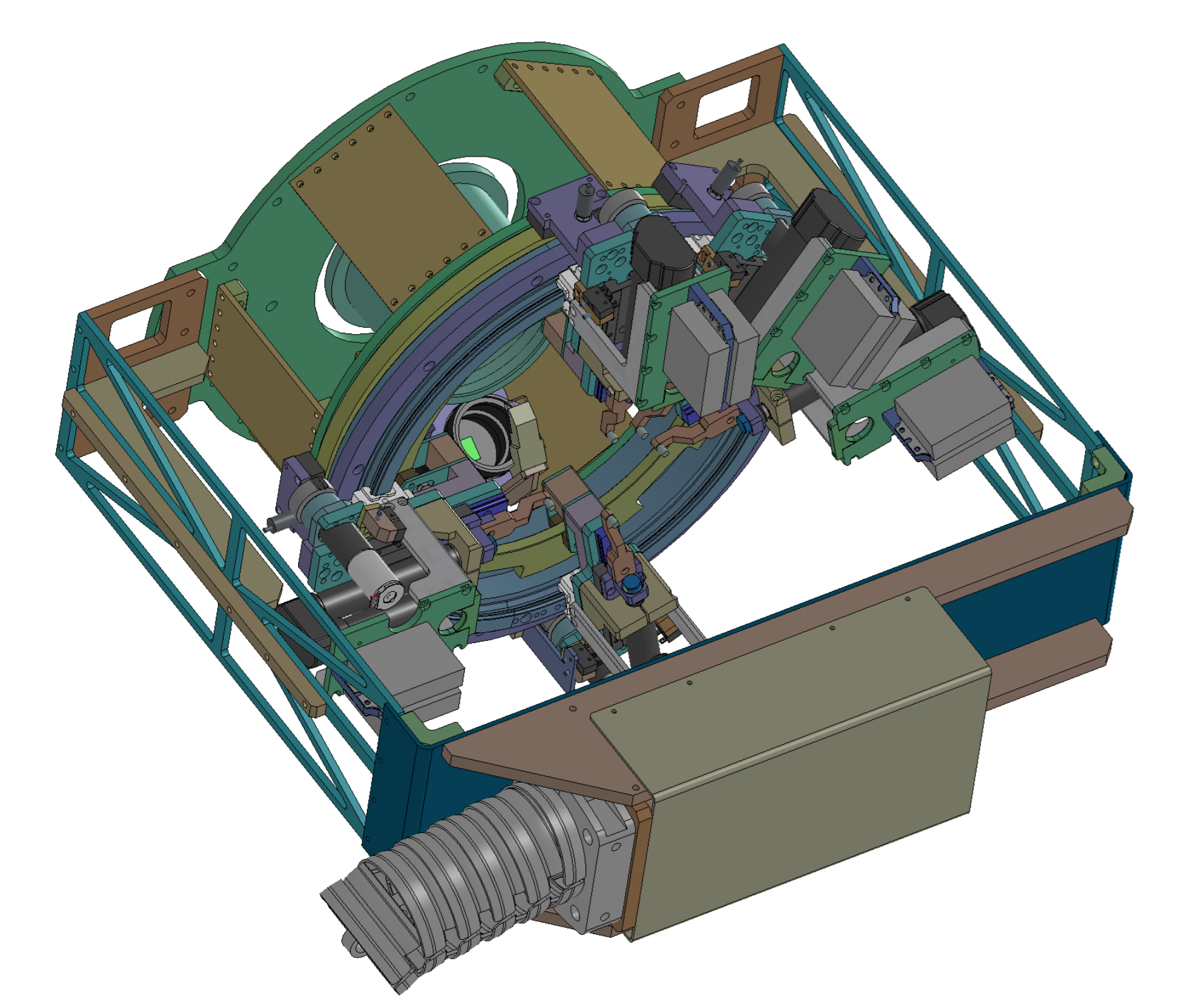}
    \end{tabular}
  \end{center}
  \caption[Focal plane positionners]{An ExTrA focal plane unit plan (left panel) and 3D implementation (right panel). The main components are: 5 field units with radial and polar movements, and an off-axis guiding camera with a dedicated pick-up mirror.\label{fig:robot}}
\end{figure}

FUs will be positionned by specific positionner system developped at IPAG. The system shown in Fig.~\ref{fig:robot} allows a polar positonning of the field units in the 80~mm field. The absolute position precision is \textless 20~$\mu$m, whereas the relative position precision is $\sim1~\mu$m.

\subsubsection{Spectrograph}

\begin{figure}
  \begin{center}
    \begin{tabular}{c}
      \hspace{0.2cm}\includegraphics[width=13.8cm]{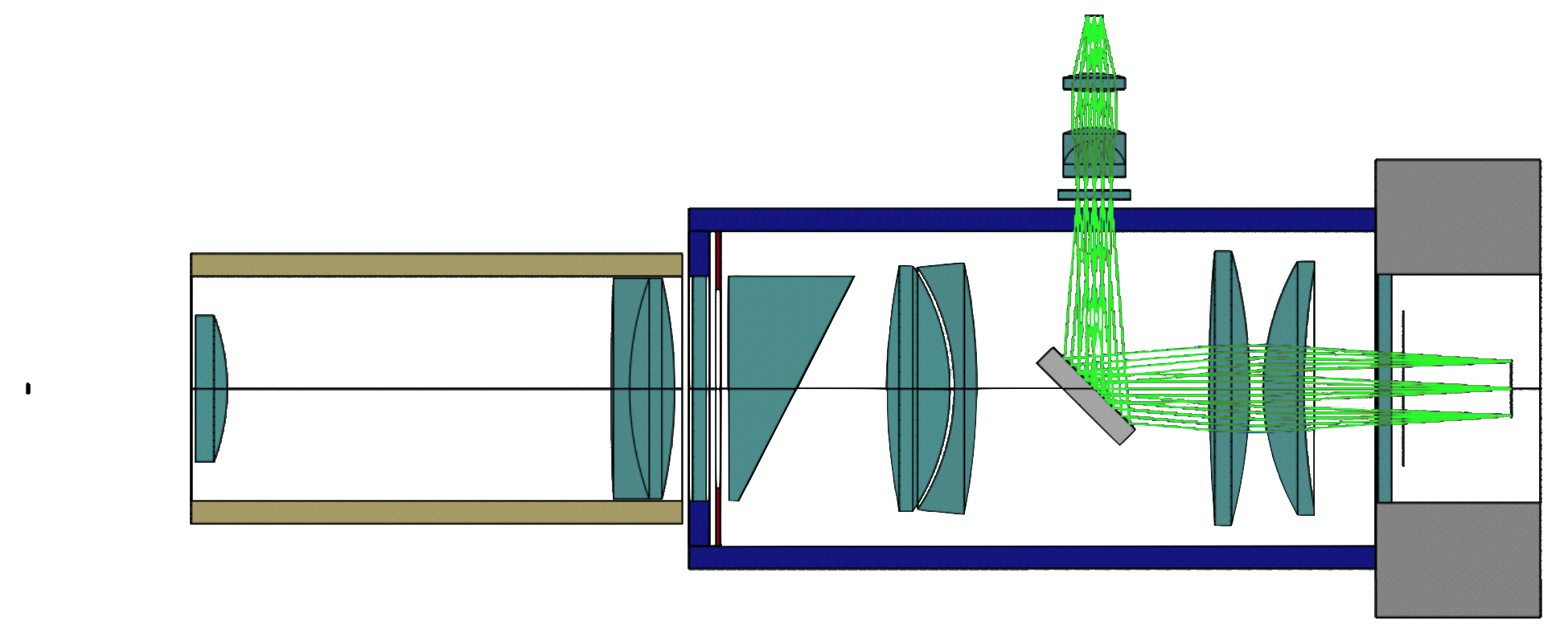}\\
      \includegraphics[width=14cm]{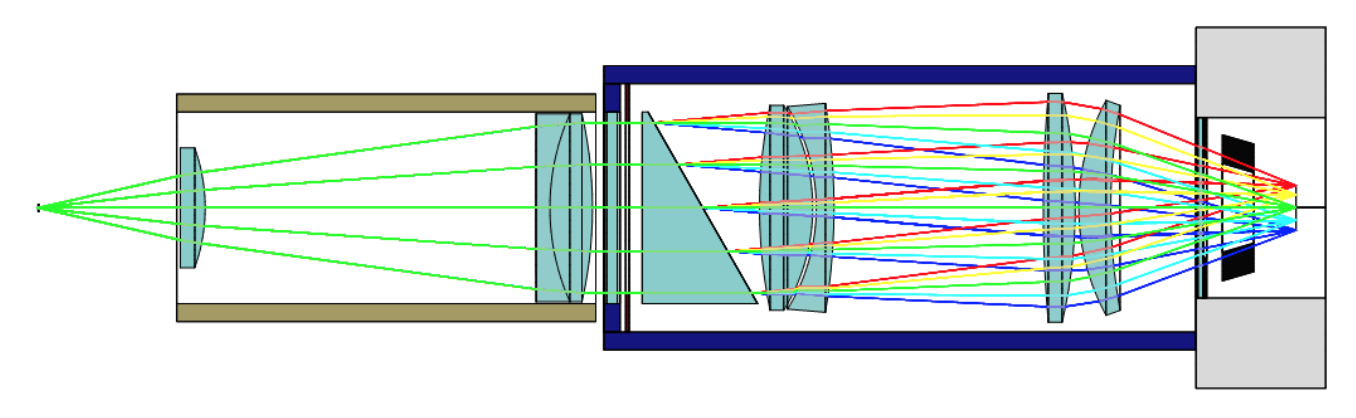}
    \end{tabular}
  \end{center}
  \caption[Spectrograph]{The two configurations of the spectrograph: centering mode (upper panel), and science mode (lower panel).\label{fig:spectrograph}}
\end{figure}

The spectrograph allows to image the fibre bundles in centering mode, and with a different light path disperse the science fibres using a prism (see Fig.~\ref{fig:spectrograph}). The centering mode is selected through a flipping mirror that gets in the science light path. There are no moving parts in the spectrograph science optical path to secure the precision and repeatability of the measurements. The science fibers are arranged in drawners (one per field unit) at the spectrograph image plane. The spectral range is from 0.85 to 1.55~$\mu$m, the resolution is R$\sim$200, and collect the light from the science fibres at F/3.5. There are 60 science fibers at the spectrograph entrance, but only 30 are used simultaneoulsy (8\arcsec\ or 4\arcsec\ aperture).

The spectrograph is located on the auxiliary building in an isolated room thermally controlled.

\subsubsection{Camera}

The camera is the NIRvana 640 LN made by Princeton Instruments around an InGaAs low-noise detector from Xenics, composed by an array of 640x512 20~$\mu$m square pixels. The detector is cooled to 83~K with liquid nitrogen. It is sensible from 0.85 to 1.55~$\mu$m, with typical quantum efficiency of $\sim$70\%. Each spectra will be dispersed on $\sim$3000 pixels in the camera.

\subsubsection{Software}

\begin{figure}
  \begin{center}
    \begin{tabular}{c}
      \hspace{0.2cm}\includegraphics[height=10cm]{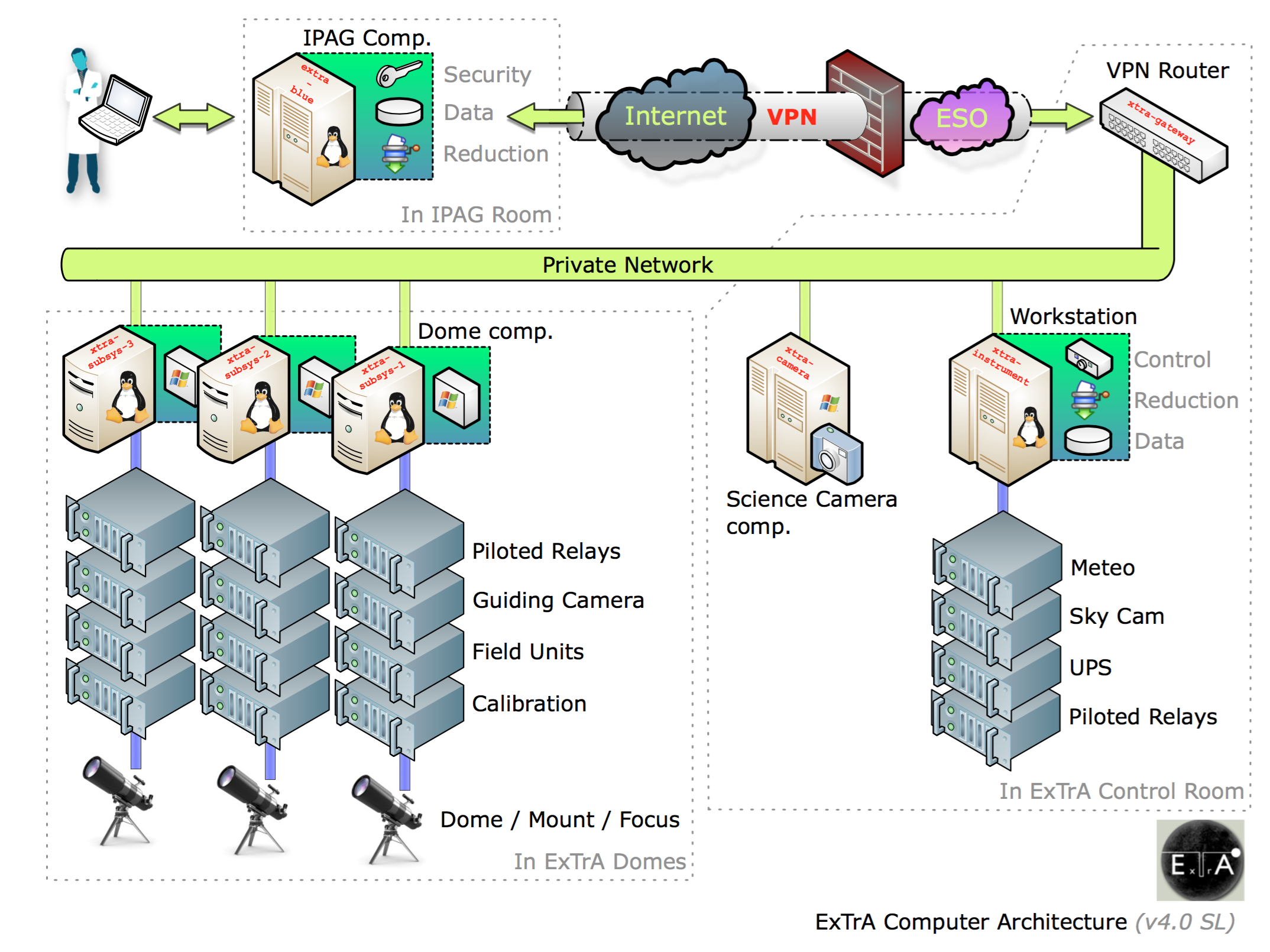}
    \end{tabular}
  \end{center}
  \caption[Software]{ExTrA computer architecture.\label{fig:software}}
\end{figure}

ExTrA will be a robotic observatory. The experiment will be controlled by a main server from the auxiliary building with TANGO\footnote{\url{http://www.tango-controls.org/}}, and Python functions. This main server will control each ExTrA subsystem through a software bus. One machine at each dome will control the telescope, mount, guiding camera, and dome, with ASCOM\footnote{\url{http://ascom-standards.org/}} through MaxIm DL\footnote{\url{http://www.cyanogen.com/maxim_main.php}}. The ExTrA computer architecture is shown in Fig.~\ref{fig:software}.

\subsubsection{Site}

\begin{figure}
  \begin{center}
    \begin{tabular}{c}
      \includegraphics[height=5cm]{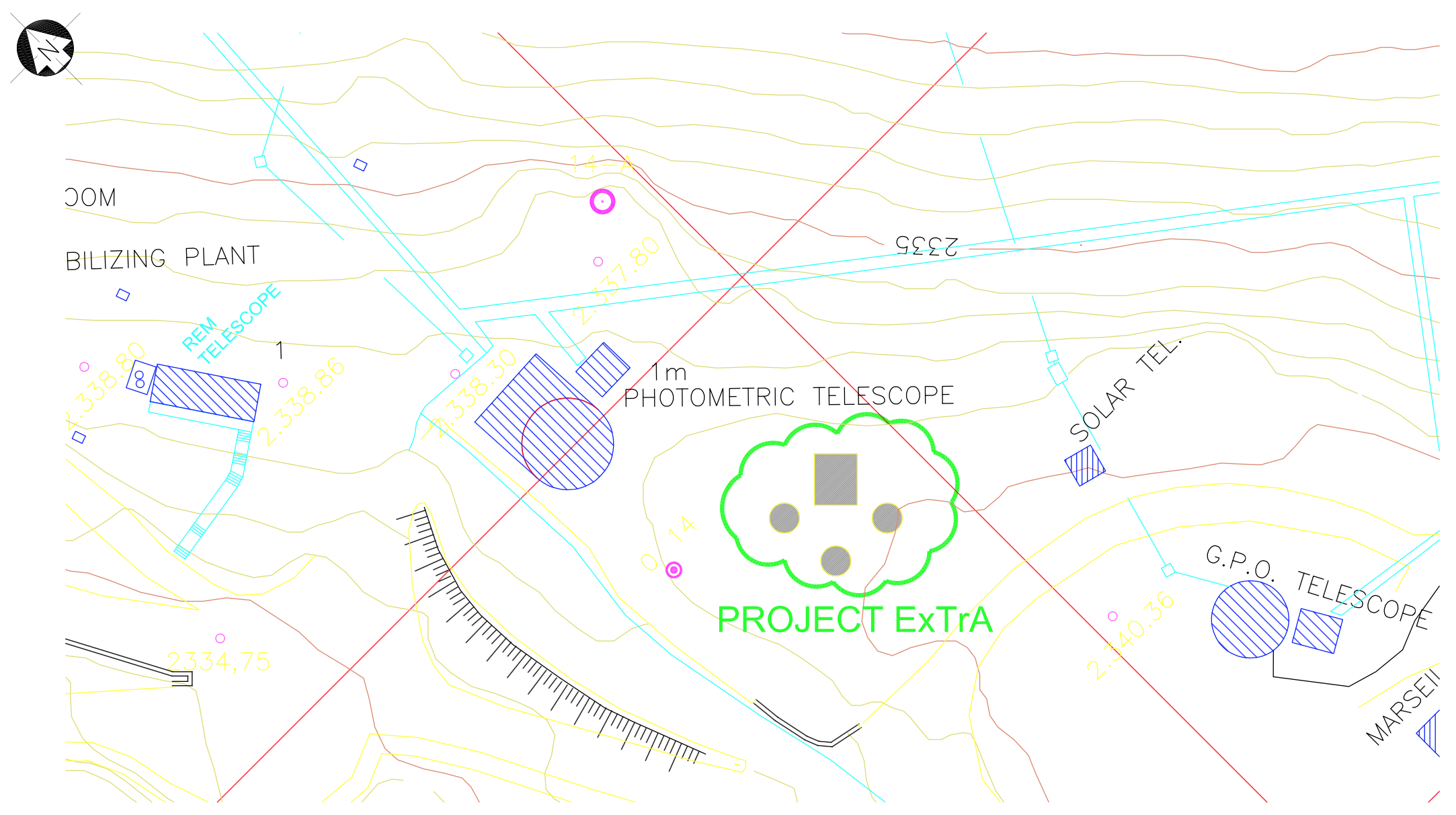}\hspace{1cm}\includegraphics[height=4.5cm]{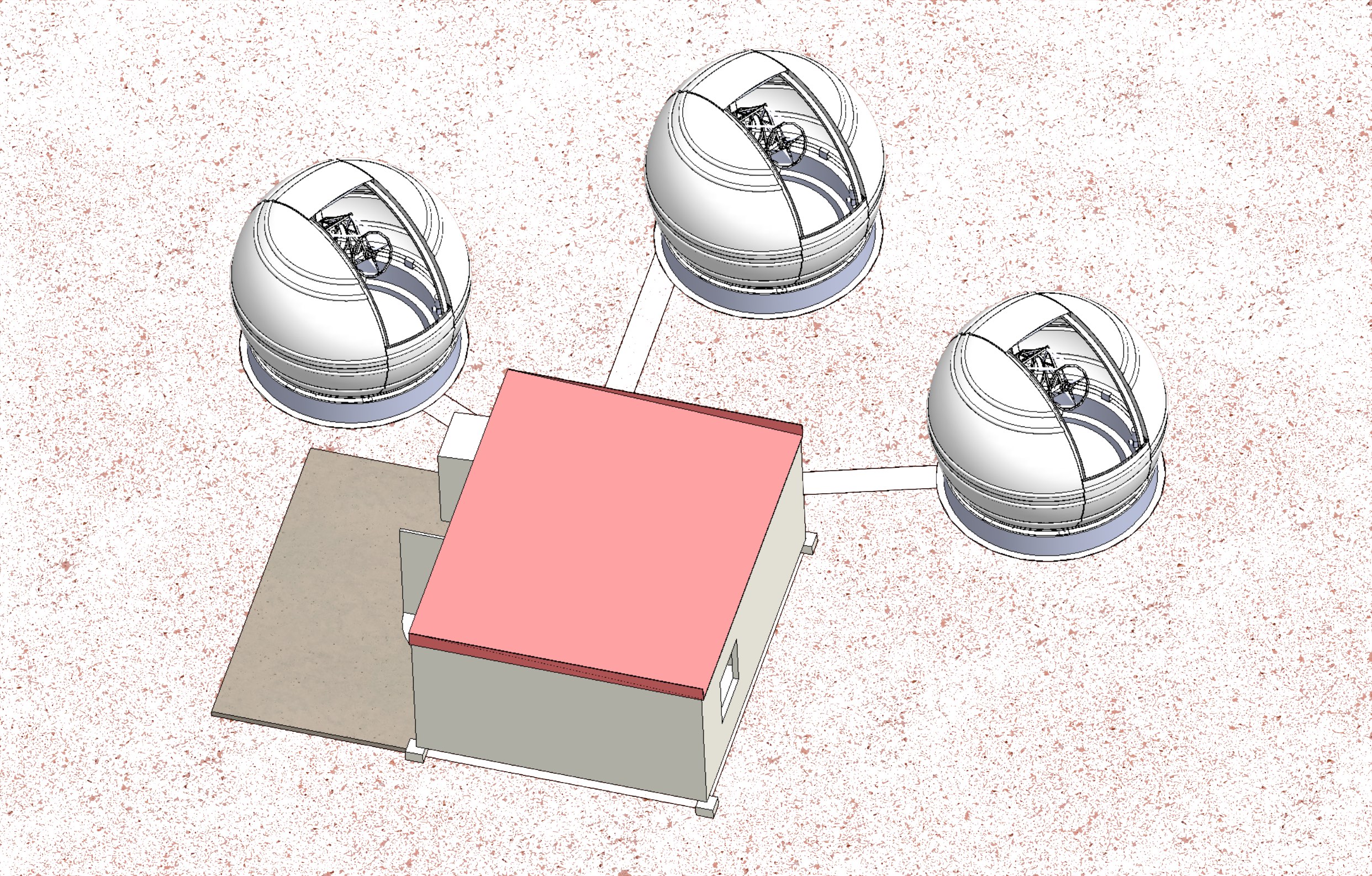}
    \end{tabular}
  \end{center}
  \caption[La Silla]{The loaction of the ExTrA facility at La Silla (left), and 3D view (right).\label{fig:site}}
\end{figure}

The ExTrA facility, will be located at La Silla observatory (see Fig.~\ref{fig:site}). It is composed by the three domes and an auxiliary building that host the spectrograph, main server, weather station, all-sky camera, and a webcam. 

\section{Performances}
\label{sec:performances}

\begin{figure}
  \begin{center}
    \begin{tabular}{c}
      \includegraphics[height=8cm]{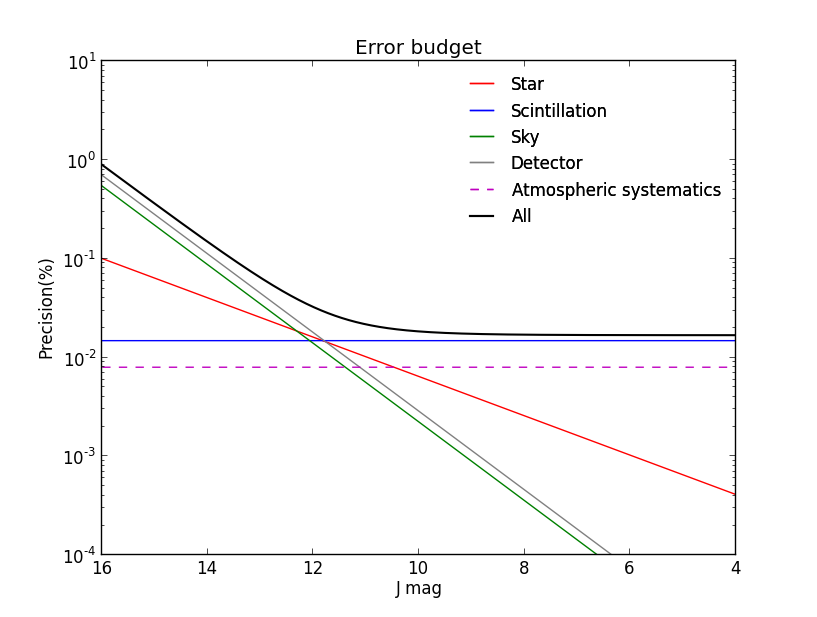}
    \end{tabular}
  \end{center}
  \caption[Error budget]{Precision as a function of J magnitude for 1h exposure time. The different noises contributions are shown with dashed lines.\label{fig:error}}
\end{figure}

To assess the global performances of ExTrA, we performed the simulations of all noise sources and made an error budget (Fig.~\ref{fig:error}). ExTrA has the capability to achieve 0.02\%.hr$^{-1/2}$ precision for stars brighter than J=10. When a lower precision is required, it shall achieve a 0.1\% precision in less than 200 s (overheads included) up to magnitude J=11.

Atmospheric scintillation remains an important limitation for bright stars (J$<$11 mag). Nevertheless, we will have the opportunity to explore two different strategies to mitigate that noise: 'lucky' photometry and 'conjugate plane' photometry. On the one hand indeed, ExTrA uses a CMOS infrared detector that can integrates while reading. With a sampling rate of about 1 read per second, one has the possibility to discard infrequent outliers registered during the most turbulent instants of an exposure. On the other hand, conjugate plane photometry proposes to image the plane conjugated with the high-altitude turbulent layer. In such an image, the effect of scintillation appears to be concentrated on a much smaller area, that can eventually be masked out or apodized. In ExTrA, we image the pupil of the telescope at the entrance of each fiber and imaging a conjugate plane instead is not much different. Also, the technique requires a mask per star and is thus well suited to a multi-object instrument.  

\section{Survey}
\label{sec:survey}

\subsection{ExTrA input catalog}

We compiled a list of M dwarfs from various sources, including the all sky LSPM catalog\cite{2005AJ....129.1483L}, stars within 10~pc as well as SDSS, DENIS and 2MASS confirmed M and L stars. We collected their infrared magnitudes in 2MASS catalog and computed their mass (M$_{\star}$), radius (R$_{\star}$) and temperature (T$_{\mathrm{eff}}$) with theoretical luminosity relationships\cite{1998A&A...337..403B}. We included the orbital period (P$_{\mathrm{HZ}}$), semi-major axis (a$_{\mathrm{HZ}}$), and transit duration (dT$_{\mathrm{HZ}}$) of a hypothetical Earth-like planet transiting the stars\cite{2009ApJ...698..519K}. We also computed the orbital period (P$_{\mathrm{HZ,OUT}}$) and the semi-major axis (a$_{\mathrm{HZ,OUT}}$) of a planet being on the outer edge of the habitable zone\cite{2007A&A...476.1373S}. We refined the selection to stars with declinations within $\pm$40\degree\ of the site latitude and J\textless16 mag and checked that the vast majority have suitable comparison stars\footnote{Note that whereas classical photometric searches need comparison stars of similar spectral types, our method instead allows choosing comparison stars of different spectral types, like field K~dwarfs.} in the 1 square degree field-of-view of our telescopes. Hopefully, this input catalog will soon be obsolete after GAIA 2nd release (early 2017).

\subsection{Planet distribution}
\label{sec:planetdist}

Several surveys have recently measured the occurrence rate of low-mass planets around M~dwarfs. Our HARPS survey of early to mid-M~dwarfs finds an 88$^{+56}_{-19}$\% occurrence rate\footnote{An occurrence rate greater than 100\% corresponds to more than one planet per star.} for super-Earths (1 \textless m sin i \textless 10 M$_{\oplus}$) with periods under 100~days, of which 41$^{+54}_{-13}$\% orbit in the Habitable Zone\cite{2013A&A...549A.109B}. For super-Earths orbiting late-K to early-M dwarfs (3600 \textless T$_{\mathrm{eff}}$ \textless 4100~K), the Kepler mission similarly indicates a high occurrence rate, with Ref.~\citenum{2012ApJS..201...15H} and Ref.~\citenum{2012ApJ...746...36G} respectively reporting 30$\pm$8\% and 36$\pm$8\%. $\mu$-lensing surveys, which probe (much) larger separations, also find a steeply rising planetary mass function towards lower mass planets, with for instance a 62$\pm$36\% occurrence rate for 5-10~M$_{\oplus}$ planets between 0.5 and 10~AU\cite{2012Natur.481..167C}. Other studies also reported high occurrence rates for small planets orbiting M~dwarfs, for the occurrence of habitable planets from Kepler\cite{2013ApJ...767...95D} (0.40 habitable planet per star), and the Mearth team\cite{2013ApJ...775...91B} for the occurrence of 2-4 R$_{\oplus}$ planets, with periods P\textless10~day (22$^{+52}_{-6}$\%). A consistent picture hence emerges, with HARPS, Kepler, Mearth and the $\mu$-lensing surveys all finding an occurrence rate for 1-10~M$_{\oplus}$ planets of 30-50\% per period decade, and 40~habitable-zone super-Earths for every 100~M~dwarfs. To estimate the yield of ExTrA' survey we assumed a planet distribution that is uniform in log of the period and normal in log of the mass.  We tuned the parameters and normalized the log-normal distribution such that it peaks at 1 M$_{\oplus}$ and matches the above occurrence rates. 

\subsection{Merit function}

We want ExTrA to provide as many habitable-zone transiting planets as possible. We thus designed a merit function that is the likelihood of finding a habitable-zone planet for a given observing time. For each target, we first expressed the probability P$_{\mathrm{IN}}$ of seeing a planet in transit at any given time (assuming the star hosts 1~planet in the habitable zone): P$_{\mathrm{IN}}$ = (R$_{\star}$/a$_{\mathrm{HZ}}$).(dT$_{\mathrm{HZ}}$/P$_{\mathrm{HZ}}$), that is the product of the geometric probability that the planet undergoes transit with the probability of observing when the transit occurs. Next, we used the simulated instrument performances to compute the exposure time T$_{\mathrm{EXP}}$ required to achieve the 5-$\sigma$ detection of a 0.5~R$_{\oplus}$ planet or, when such a signal is below the noise floor, to achieve a precision of 0.01\% (i.e. a 5-$\sigma$ precision of 0.05\%; see Fig.~\ref{fig:spaceparam}). Then, the merit function is: M = P$_{\mathrm{IN}}$/T$_{\mathrm{EXP}}$. Before starting operations, we plan to alter our merit function to reflect the prior  knowledge on the stellar inclination (and thus, to some extend, the transit probability) brought by the K2 mission and, in the future, by the TESS mission.

\subsection{Observing strategy}

\begin{figure}
  \begin{center}
    \begin{tabular}{c}
      \includegraphics[height=6.4cm]{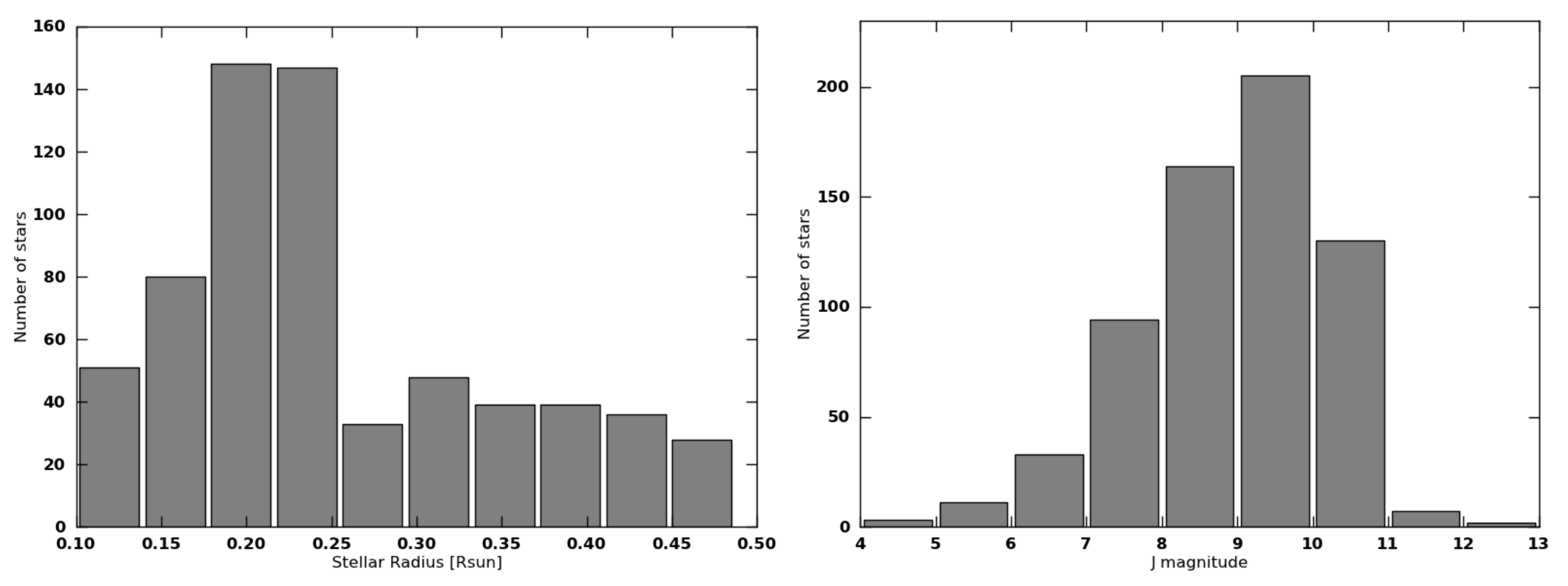}
    \end{tabular}
  \end{center}
  \caption[ExTrA sample]{Sample distributions. Histograms of stellar radii (left panel) and J magnitudes (right panel) for the 797~M and 3~L~dwarfs with the highest merit values to search for transiting habitable-zone planets.
\label{fig:sample}}
\end{figure}

ExTrA will follow the strategy proposed by the MEarth survey\cite{2009IAUS..253...37I}. Each telescope cycles on a set of N~stars, recording sparsely sampled light curves until detection of a 5-$\sigma$ drop in flux. On that trigger, the telescope stops observing other stars and records the transit egress with dense sampling. With M~telescopes, this strategy searches transits around $\sim$NxM~stars. More precisely, a Queue Scheduler prepares the observations before the night starts. It gives priority to stars with higher merit values and adapts the cadence of observations according to the possible transit durations. The required cadence equals to dT/4, which corresponds to 4 visits for a central transit and at least 1 visit for 99.5\% of all possible transit configurations. When a star starts being observed, ExTrA searches for the transiting planets with the shortest periods. They have short transit time durations and the cadence of observation is high. As longer periods are covered, the possible transits have progressively longer durations and the cadence can be slowed down. It is also required to observe the same stars on consecutive nights, until the orbital period corresponding to the outer edge of the habitable zone is covered (P$_{\mathrm{HZ,OUT}}$). Assuming a 4-year survey, ExTrA has the capability to observe 800~stars, which distributions in mass and J~magnitude are shown in Fig.~\ref{fig:sample}.

\subsection{Overheads}

The sequence of acquisition is as follow: the telescope slews to the target field (pointing precision=6\arcsec). The guiding camera image the guiding star (time=5~s, for V$\leq$12). A pointing correction is computed (time=5~s). The positionners slew to the main target and the 4~comparison stars (time=5~s, precision is now \textless1\arcsec). An exposure is done with the centering button (10~s, for J$\leq$11). The photocenter is measured with the fiber bundle. Positionners correct their position (10~s, precision \textless0.1\arcsec) and the science exposure starts. Overheads to preset a science acquisition are thus less than a minute.

\subsection{Yields}

\begin{figure}
  \begin{center}
    \begin{tabular}{c}
      \includegraphics[height=8cm]{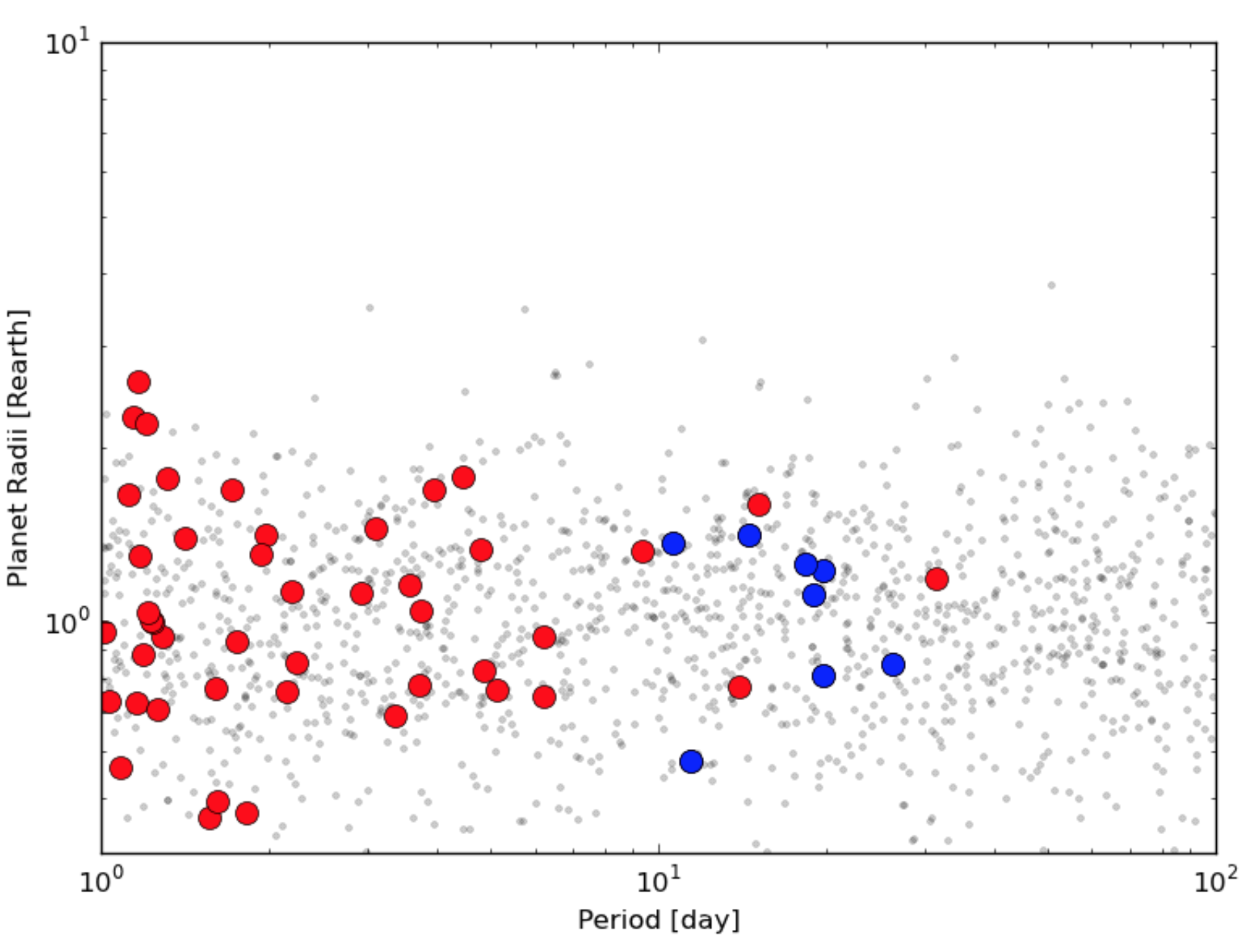}
    \end{tabular}
  \end{center}
  \caption[Yields]{Simulated yield of planet detections. Eight hundred targets were selected according to their merit of finding habitable-zone transiting planets. The planet population was generated assuming a uniform distribution in $\log$(Period) and a normal distribution in $\log$(Mass). We choose a normal distribution that peaks at 1~M$_{\oplus}$ (and an Earth-density mass-radius relationship). Note that more detections are expected if the true distribution peaks at \textless 1~M$_{\oplus}$. The planets generated are shown as gray dots, transits are red filled circles, and habitable-zone transits are blue filled circles.
\label{fig:yield}}
\end{figure}

To estimate the yield of our survey, we used the planet distribution and a simple Earth-density mass-radius relation to generate a virtual population of transiting planets (orbital periods, transit depths and transit durations), which are randomly attributed to the stars in the input catalog. We then assumed a survey of 4~years, nights of 8~hours, and an 80\% duty cycle (20\% being lost to weather). For each star in the input catalog, we computed the number of measurements N$_{\mathrm{MES}}$ obtained when the star has been observed for a period P$_{\mathrm{HZ,OUT}}$ as well as the total observing time T$_{\mathrm{OBS}}$ dedicated to these measurements. Then, we looped over the 800~stars with the highest merit. When the star has one or more virtual planets, we computed the number of measurements N$_{\mathrm{DET}}$ taken during transit events of each planet and considered the planet detected if N$_{\mathrm{DET}}$ \textgreater 1. Eventually, we found that the ExTrA survey shall detect $\sim$50~transiting planets, including \textgreater 5 in the habitable zone (see Fig.~\ref{fig:yield} for a favorable realization).

\subsection{Magnetic activity}

Low-mass stars are notoriously active, with spots, plages and flares modulating their luminosity, which introduce an astrophysical noise to photometric observations. The flux variations caused by spots and plages occur on similar time-scale as the stellar rotation, which is of the order of days, and therefore much different than the few hours timescale of transit events. Flares can occur on shorter time-scales but are relatively rare events with characteristic flux increases. They can thus be identified and rejected easily.

\subsection{False positives and mass measurements with radial-velocity}

Potential planets detected in light curves need follow-up. Unlike most photometric transit searches, surveys targeting nearby M dwarfs have no false positives from blended background binaries. Indeed, nearby stars have large proper motions and using past sky, we will eliminate M dwarfs now blended with a background object. The follow-up will thus directly use spectrographs to reject blends from bound companions to the target, and determine the mass of the planet. 
    After the Mearth team detected the transiting planet GJ1214b in their light curves, we measured its mass with the HARPS spectrograph. The faintest stars of ExTrA survey are only $\sim$2~mag fainter than GJ1214 and would thus be amenable to follow-up as well. Much lower-mass detections are however expected from ExTrA. When found around fainter stars, they shall be at the limit of HARPS capability. Fortunately, ESPRESSO and SPIRou, two HARPS-fashioned spectrographs, are on schedule to be operational from 2016-2017 on. Compared to HARPS, ESPRESSO will be more stable, have extended wavelength coverage in the red and, more importantly, could be (fiber-)fed by up to four 8-m telescopes (simultaneously). SPIRou will transpose the HARPS m/s precision to the infrared, making radial-velocity measurement of M dwarfs extremely efficient. Both ESPRESSO and SPIRou will thus provide the possibility to measure the mass of all ExTrA detections within the duration of our project.

\section{Conclusion}

This paper presented the project ExTrA and described the implementation of a new facility to search for extra-solar planets. ExTrA offers a novel method to improve on the precision of ground-based photometry. The method makes use of a multi-object spectrograph to add spectroscopic resolution and correct the atmospheric variations that would otherwise hinder ground-based observations. We are implementing the ExTrA method to detect new exoplanets with three small-size telescopes (60 cm in diameter) equipped with a multi-object spectrograph (fiber-fed, R\textgreater200, $\lambda$=0.8-1.6~$\mu$m) to perform differential spectro-photometry on a sample of $\sim$800~M~dwarfs. Besides running its own survey, ExTrA will be a valuable resources to follow-up M dwarfs with candidates found by both K2 and TESS mission.

%%%%%%%%%%%%%%%%%%%%%%%%%%%%%%%%%%%%%%%%%%%%%%%%%%%%%%%%%%%%%
\acknowledgments     %>>>> equivalent to \section*{ACKNOWLEDGMENTS}       
 
The ExTrA team acknowledges funding from the European Research Council under the ERC Grant Agreement n. 337591-ExTrA and support of the French Agence Nationale de la Recherche (ANR), under program ANR-12-BS05-0012 Exo-Atmos.

%%%%%%%%%%%%%%%%%%%%%%%%%%%%%%%%%%%%%%%%%%%%%%%%%%%%%%%%%%%%%
%%%%% References %%%%%

\bibliography{extra}   %>>>> bibliography data in report.bib
\bibliographystyle{spiebib}   %>>>> makes bibtex use spiebib.bst

\end{document}